\documentclass[cpp,fleqn]{w-art}
\usepackage{times}
\usepackage{w-thm}
\usepackage[utf8]{inputenc}
\usepackage{amsmath}
\usepackage{amssymb}
\usepackage{graphicx}
\usepackage{sidecap}
\usepackage[english]{babel}
\begin{document}
\DOIsuffix{theDOIsuffix}
%%
%% issueinfo for header and copyright line
\Volume{42}
\Issue{1}
\Month{01}
\Year{2013}
%%
%%    First and last pagenumber of the article. If the option
%%    'autolastpage' is set (default) the second argument may be left empty.
\pagespan{1}{}
%%
%%    Dates will be filled in by the publisher. The 'reviseddate' and
%%    'dateposted' (Published online) entry may be left empty.
\Receiveddate{}
\Reviseddate{}
\Accepteddate{}
\Dateposted{}

%% \pretitle{Editor's Choice}

%% We have a short and a long form for the title. The short form
%% (optional argument) goes into the running head.

\title[Nonequilibrium Green functions approach to diffusion at strong coupling]{Toward a Nonequilibrium Green functions approach to diffusion in strongly coupled finite quantum systems}

%% Please do not enter footnotes or \inst{}-notes into the optional
%% argument of the author command. The optional argument will go into
%% the header.  If there is only one address the marker \inst{x} may be
%% omitted.

%% Information for the first author.
\author[M. Bonitz]{M. Bonitz\footnote{Corresponding
     author: e-mail: {\sf bonitz@physik.uni-kiel.de}, Phone: +49\,431\,8804122,
     Fax: +49\,431\,8804094}}
\author[N. Schl\"unzen]{N. Schl\"unzen}
\author[S. Hermanns]{S. Hermanns}
\address[]{Christian-Albrechts-Universität zu Kiel, Institut für Theoretische Physik und Astrophysik, Leibnizstraße 15, 24098 Kiel, Germany}
\newcommand{\todo}[1]{\textcolor{red}{\underline{#1}}}
\newcommand{\eq}[1]{Eq.~(\ref{#1})}
\newcommand{\nn}{\nonumber}
\newcommand{\e}[1]{\mathrm{e}^{#1}}
\newcommand{\op}[1]{\hat{#1}}
\newcommand{\bra}[1]{\langle{#1}|}
\newcommand{\ket}[1]{|{#1}\rangle}
\newcommand{\braket}[2]{\langle{#1}|{#2}\rangle}
\newcommand{\eqsand}[2]{Eqs.~(\ref{#1}) and~(\ref{#2})}
\newcommand{\cc}{{\cal C}}
\newcommand{\mean}[1]{\langle #1 \rangle}
\newcommand{\tc}{T_{\cal C}}
\newcommand{\dc}{\delta_{\cal C}}
\newcommand{\ii}{\mathrm{i}}
\newcommand{\thc}{\theta_{\cal C}}
\newcommand{\intc}[1]{\int_{\cal C}\mathrm{d}{#1}\;}
\newcommand{\intlim}[3]{\int_{#1}^{#2}\mathrm{d}{#3}\;}
\newcommand{\mret}{{\mathrm{ret}}}
\newcommand{\madv}{{\mathrm{adv}}}

%%    \dedicatory{This is a dedicatory.}
\begin{abstract}
Transport properties of strongly correlated quantum systems are of central interest in condensed matter, ultracold atoms and in dense plasmas. There, the proper treatment of strong correlations poses a great challenge to theory. Here, we apply a Nonequilibrium Green Functions approach using a lattice model as a basic system. This allow us to treat a finite spatially inhomogeneous system with an arbitrary nonequilibrium initial state. Placing all particles initially to one side of the system allows for a nonequilibrium study of diffusion. Strong correlation effects are incorporated via T-matrix selfenergies.
\end{abstract}
%% maketitle must follow the abstract.
\maketitle                   % Produces the title.

%% \tableofcontents  % Produces the table of contents.
\section{Introduction}\label{s:intro}
Particle and energy transport of strongly correlated quantum systems are of growing current interest in condensed matter \cite{pavarini11,balzer_prb09, uimonen11}, ultracold quantum gases \cite{schneider_np12,kajala_prl11,ronzheimer13}, in dense plasmas and warm dense matter \cite{rudd_14,bonitz_cpp99}. 
The proper theoretical description of the strong correlations occurring within these systems is highly demanding. For classical systems, rigorous results can be obtained from molecular dynamics simulations of transport, e.g., \cite{daligault_prl06,ott_prl09}. Also, progress in analytical models has been achieved recently \cite{baalrud_pop14,kaehlert_pre14}. The situation is essentially more complex with quantum systems due to the absence of first-principle transport simulations. One approach to transport theory is given by linear response theory around an equilibrium state or via computation of equilibrium fluctuations, e.g., via ab-initio molecular dynamics \cite{karasiev_prb13}. These density functional theory based methods face the familiar problem with the treatment of strong correlation effects. An alternative that also treats systems far from equilibrium are exact diagonalization (CI) methods or multiconfiguration time-dependent Hartree Fock (MCTDHF), e.g., \cite{hochstuhl_jcp11} and references therein, but their effort grows exponentially with the system size, and they are thus limited to small particle numbers on the order of $15$.
To overcome these limitations, here, we apply a Nonequilibrium Green Functions (NEGF) approach using a lattice model as a basic system. This allows us to accurately treat a finite spatially inhomogeneous system with an arbitrary nonequilibrium initial state. In particular, placing all particles initially to one side of the system allows for a nonequilibrium study of diffusion. 

NEGF simulations have seen a rapid progress in recent years and have been applied to dense plasmas \cite{bonitz_cpp99}, electron-hole plasmas in semiconductors \cite{bonitz_jpcm96, kwong98, gartner06}, quantum transport \cite{puigvonfriesen09, uimonen11}, nuclear matter and high energy physics \cite{rios11, garny11}. Recently, applications to inhomogeneous systems have been performed including electrons in atoms \cite{dahlen07, balzer_pra10} and the Hubbard model, e.g., \cite{puigvonfriesen09,puigvonfriesen10,bonitz_cpp13}. Lattice models have the advantage that complicated approximations for the selfenergy may be treated accurately. This was shown for the T-matrix selfenergy in Refs.~\cite{puigvonfriesen09,puigvonfriesen10}, although the simulations had to be restricted to short propagation times, due to the large computational effort. We have recently shown that these limitations can be overcome in part by employing the generalized Kadanoff Baym ansatz (GKBA) \cite{lipavsky86}. Within the second Born approximation, we could achieve long simulation times and an improved long time behavior compared to full two-time simulations, e.g., \cite{hermanns_prb14}. Here, we extend this approach to strong coupling by performing T-matrix calculations in combination with the GKBA. It is the purpose of this paper to show the first results of this approximation and apply them to a study of diffusion and its dependence on the strength of correlations.

\section{Theory}
\subsection{Nonequilibrium diffusion in a correlated Hubbard model}\label{ss:hubbard}
%
%\section{Nonequilibrium diffusion in a correlated Hubbard model}\label{s:hubbard}
%
We are interested in the dynamical behavior of a finite quantum system beyond the regime of linear response. Strong correlation effects should be accounted for and we want to analyze the full nonequilibrium dynamics. This ambitious set of goals can be realized for the Hubbard model with hopping amplitude $J$ and on-site interaction $U$.
The initial hamiltonian, for times $t<0$, reads
\begin{align}
\label{eq:h0}
 \op{H}_{0}=-J\sum_{<s,s'>}\sum_{\sigma=\uparrow,\downarrow}\op{c}^\dagger_{s,\sigma}\op{c}_{s',\sigma}+U\sum_{s}\op{n}_{s,\uparrow}\,\op{n}_{s,\downarrow}\;,
\end{align}
where $s$ and $s'$ label the discrete sites, and $<\!s,s'\!>$ indicates nearest-neighbor sites, and the total number of sites is denoted $N_s$. Further, $\op{n}_{s,\sigma}=\op{c}^\dagger_{s,\sigma}\op{c}_{s,\sigma}$ denotes the density operator, and the energy (time) is measured in units of $J$ (the inverse hopping amplitude $J^{-1}$). Below, we use the hopping amplitude $J$ as the energy unit, thus the interaction strength $U$ will be measured in units of $J$.
In this first analysis, we will concentrate on a 1D chain containing $N_s=18$ sites.
To initiate diffusion, we choose an initial state ($t < 0$) where all $N$ particles are confined to the leftmost 4 sites. 
The particle number will be varied from $N=2$ to $N=8$ which allows one to see the influence of density (filling) effects. At time $t = 0$, the system is strongly perturbed by removal of the (virtual) barrier between sites four and five, and particle transport sets in. We do not use periodic boundary conditions, so the dynamics is influenced by interferences arising from reflections at the left wall whereas reflections from the right wall are practically avoided by properly limiting the simulation time.

\subsection{Nonequilibrium Green Functions. T-matrix approximation}\label{s:negf}
To describe the electron dynamics in the Hubbard model, the central quantity is the one-particle nonequilibrium Green function defined on the complex Keldysh contour $\cc$ ($\tc$ denotes contour ordering of the times $t$ and $t'$), e.g., \cite{balzer_lnp13},
\begin{align}
\label{gdef}
 g_{ss'}^\sigma(t,t')&=-\ii\mean{\tc\,\op{c}_{s,\sigma}(t)\,\op{c}^\dagger_{s',\sigma}(t')},
%\\
% &=\thc(t-t')\,g_{ss'}^{\sigma,>}(t,t')+\thc(t'-t)\,g_{ss'}^{\sigma,<}(t,t')\;,\nn
\end{align}
with the lattice site indices $s$, $s'$ and the spin projection $\sigma \in \uparrow,\downarrow$. Here and below, we use atomic units with $\hbar=1$, and $\mean{ \ldots}$ means averaging in the grand canonical ensemble.
From the NEGF, all relevant observables can be computed \cite{balzer_lnp13}, including the density matrix, $\rho_{ss'}^{\sigma}(t)=-\ii g_{ss'}^{\sigma,<}(t,t)$, and the time-dependent spin density on site $s$, $\rho_{ss}^{\sigma}(t)=\langle \op{n}_{s,\sigma}(t) \rangle \equiv n_{s,\sigma}(t)$. 
%or equivalently $\rho_{ss'}^{\sigma,>}(t)=-\ii g_{ss'}^{\sigma,>}(t,t)$.
%
%Due to the two-time dependence of the Green function, a systematic treatment of dynamic correlation effects is possible where interactions, quantum and spin effects as well as coupling to a (possibly strong) external field is properly taken into account. 
The NEGF formalism provides the basis for a selfconsistent treatment of quantum, spin and correlation effects, thereby
maintaining the conservation laws \cite{book_kadanoffbaym_qsm,hermanns_prb14}. Moreover, it allows for a systematic construction of approximations via Feynman diagrams.
The equations of motion for the NEGF (\ref{gdef}) are the Keldysh-Kadanoff-Baym equations (KBE) and their adjoints~\cite{book_kadanoffbaym_qsm,balzer_lnp13},
\begin{equation}
\label{kbe}
 \left(-\ii\frac{\partial}{\partial t}\delta_{s\bar{s}}-h^\sigma_{s\bar{s}}(t)\right)g_{\bar{s}s'}^\sigma(t,t')=\dc(t-t')\delta_{ss'}+\intc{\bar{t}}\Sigma^\sigma_{s\bar{s}}(t,\bar{t})g_{\bar{s}s'}^\sigma(\bar{t},t')\;,
\end{equation}
where summation over the repeated site index $\bar{s}$ is implied on the left and right sides. On the l.h.s., $h^\sigma_{ss'}$ is the matrix of single-particle energy contributions [arising from the tunneling part of the hamiltonian (\ref{eq:h0})], whereas pair interactions [contributions proportional to $U$ in (\ref{eq:h0})] are accounted for by the one-particle self-energy $\Sigma^\sigma_{ss'}(t,t')$ [it includes a time-diagonal part---the Hartree-Fock selfenergy---and a time non-local ``correlation'' part $\Sigma^{\textnormal{cor}, \sigma}_{ss'}(t,t')$]. 
%Note that, in Eq.~(\ref{kbe}), the time arguments are defined on the Keldysh contour meaning that the functions $g$ and $\Sigma$ possess an internal matrix structure depending on how the time arguments are positioned on the contour $\cc$, for details see Ref.~\cite{balzer_lnp13}. Of particular importance are the two-time correlation functions $g_{ss'}^{\sigma,\gtrless}(t,t')$ that determine all relevant time-dependent observables, as discussed above.

%The KBE are (formally) closed equations for the Green functions but they require knowledge of the selfenergy. 
%For this, various approximation schemes and Feynman diagram methods can be used. 
The simplest approximation is the Hartree-Fock (HF) approximation where correlations are neglected entirely ($\Sigma^{\textnormal{cor}, \sigma}_{ss'}\equiv 0$). It is commonly expected that this is a reasonable approximation for weak coupling, $U\ll 1$. However, it has been shown in Ref.~\cite{hermanns_prb14} that, even for small $U$, in nonequilibrium situations correlation effects may play a crucial role, in particular, for the long-time behavior. Thus, a proper many-body description requires to include correlations. The lowest order for the correlation selfenergy is given by the second order Born (2B) approximation. The relaxation dynamics of finite Hubbard clusters in nonequilibrium revealed \cite{puigvonfriesen10,hermanns_prb14,lacroix14} that the Born approximation works well for weak coupling, $U \lesssim 0.1$. For larger coupling, the simulations have a limited time range where they are valid that shrinks as  $U^{-1}$.
However, in many quantum systems the coupling parameter exceeds one, i.e. $U>1$, in the present model.  Examples are solid state lattice systems, warm dense matter or cold atomic gases. 
%This leads to a large number of interesting physical effects such as formation of stable fermion pairs (``doublons''), e.g. \cite{schneider_np12,kajala_prl11}. 
To capture this kind of physics with NEGF requires to use selfenergies that sum the whole Born series. The proper approximation is the T-matrix selfenergy which is given by (on the Keldysh contour)
\begin{align}
 \Sigma^{\textnormal{cor}}_{ik}(t_1,t_2) &= \ii T_{ik}(t_1,t_2)  G_{ki}(t_2,t_1), 
\label{eq:sigma_t}\\
 T_{ik}(t_1,t_2) &= \pm \ii U^2\, G^\textnormal{H}_{ik}(t_1,t_2) + 
\ii U \intc{\bar{t}} G^{\textnormal{H}}_{il}(t_1, {\bar t})T_{lk}({\bar t},t_2),
\label{eq:lsg}\\
G^\textnormal{H}_{ik}(t_1,t_2) &= G^{\uparrow}_{ik}(t_1,t_2)\, G^{\downarrow}_{ik}(t_1,t_2),
\end{align}
where the T-matrix $T$ constitutes an effective interaction that is determined by the Lippmann-Schwinger equation (\ref{eq:lsg}), e.g. \cite{book_kadanoffbaym_qsm,kremp_ap97,semkat_jmp00} and, in the weak coupling limit, reduces to the second Born approximation (first term); the ``+'' (``-'') in Eq.~(\ref{eq:lsg}) refers to bosons (fermions) and summation over $l$ is implied. The use of this complex approximation under full nonequilibrium conditions has only recently become possible for the Hubbard model, e.g. \cite{puigvonfriesen09,hermanns_prb14}. Comparisons with exact diagonalization calculations (CI) confirmed the high accuracy of this approximation \cite{puigvonfriesen10}.
 
\subsection{Time propagation using the generalized Kadanoff-Baym ansatz}\label{ss:gkba}
After preparing a correlated initial state, e.g., \cite{semkat_jmp00,stefanucci13}, the system (\ref{kbe}) is propagated in the two-time plane by computing the NEGF as a function of both time arguments. Due to the time-memory structure of the collision integral in Eq.~(\ref{kbe}), the NEGF at all times and for all values of the site and spin indices has to be stored in memory leading to substantial memory and CPU time requirements. Here, substantial advances could be recently achieved via an optimized program structure and parallelization~\cite{balzer_pra10}. 
To relax these limitation, we have recently developed solutions of the KBE in the single-time limit.
This is achieved by applying the generalized Kadanoff Baym ansatz (GKBA)~\cite{lipavsky86,hermanns12_pysscripta,hermanns_prb14}, where the two-time functions appearing in the collision integral of Eq.~(\ref{kbe}) are ``reconstructed'' from their values on the time-diagonal according to
\begin{align}
\label{gkbadef}
 g_{ss'}^{\sigma,\gtrless}(t,t')=-g^{\sigma,\mret}_{s\bar{s}}(t,t')\,\rho_{\bar{s}s'}^{\sigma,\gtrless}(t')+\rho_{s\bar{s}}^{\sigma,\gtrless}(t)\,g^{\sigma,\madv}_{\bar{s}s'}(t,t')\;.
\end{align}
Here, $g^{\gtrless}$ and $g^{\mret/\madv}$ are the correlation and retarded/advanced components, respectively, of the 
Keldysh matrix function $G$ appearing in the equations above, for details see \cite{book_kadanoffbaym_qsm, balzer_lnp13}.
In Eq.~(\ref{gkbadef}), summation over $\bar{s}$ is implied and we denoted $\rho_{\bar{s}s'}^{\sigma, <}(t)=\rho_{\bar{s}s'}^{\sigma}(t)$, and $\rho_{\bar{s}s'}^{\sigma, >}(t)=1 \pm \rho_{\bar{s}s'}^{\sigma}(t)$,  where ``+'' (``-'') again refers to bosons (fermions). Finally, the two-time retarded and advanced propagators $g^{\sigma,\mret}_{ss'}(t,t')$ and $g^{\sigma,\madv}_{ss'}(t,t')$ are computed in Hartree-Fock approximation (this approximation will be called HF-GKBA)
\begin{align}\label{eq:hf_prop}
 g^{\sigma,\mret/\madv}_{ss'}(t,t')=\left.\mp\ii\thc(\pm[t-t'])\exp{\left(-\ii\intlim{t'}{t}{\bar{t}}h^\sigma_\mathrm{HF}(\bar{t})\right)}\right|_{ss'}\;,
\end{align}
where $h^\sigma_\mathrm{HF}(t)$ denotes the single-particle time-dependent Hartree-Fock hamiltonian.
%
%We emphasize that the reconstruction of the greater and lesser components of the Green function with real $t$ and $t'$ is sufficient as long as the method of adiabatic switching is applied to generate the correlated initial (ground) state by time propagation, for details see \cite{hermanns12_pysscripta}. 
The quality of the GKBA has been tested before for macroscopic spatially homogeneous systems \cite{bonitz_jpcm96, bonitz_pla96}. In contrast to damped propagators \cite{bonitz_epjb99}, the present HF-GKBA has been found to be 
total energy conserving \cite{hermanns_prb14} and to reduce artificial damping problems of full two-time simulations of strongly driven finite Hubbard systems \cite{puigvonfriesen10}.

%In general, the application of MBAs and other simplifications such as the GKBA require thorough justification as they can lead to unphysical effects such as self-interaction errors and spurious dynamical excitations~\cite{hermanns12_prb}, bistability~\cite{khosravi12} or artificial steady states~\cite{puigvonfriesen10}. For this reason, we, in this contribution, extend previous work on weak perturbations~\cite{balzer12_epl}, which included the discussion of double excitations (see also~\cite{sakkinen12}), to the nonlinear regime and analyze the performance of the GKBA for a finite system with the two-body interactions being treated in the second Born approximation.

%
\sidecaptionvpos{figure}{t}
\begin{SCfigure}[\sidecaptionrelwidth][h]
\caption{Initial density for $N=2, 4, 6, 8$ particles for weak (circles) and strong (triangles) coupling in Hartree-Fock approximation. Sites $5 \dots 18$ are empty initially.}
\includegraphics[width=0.6\textwidth]{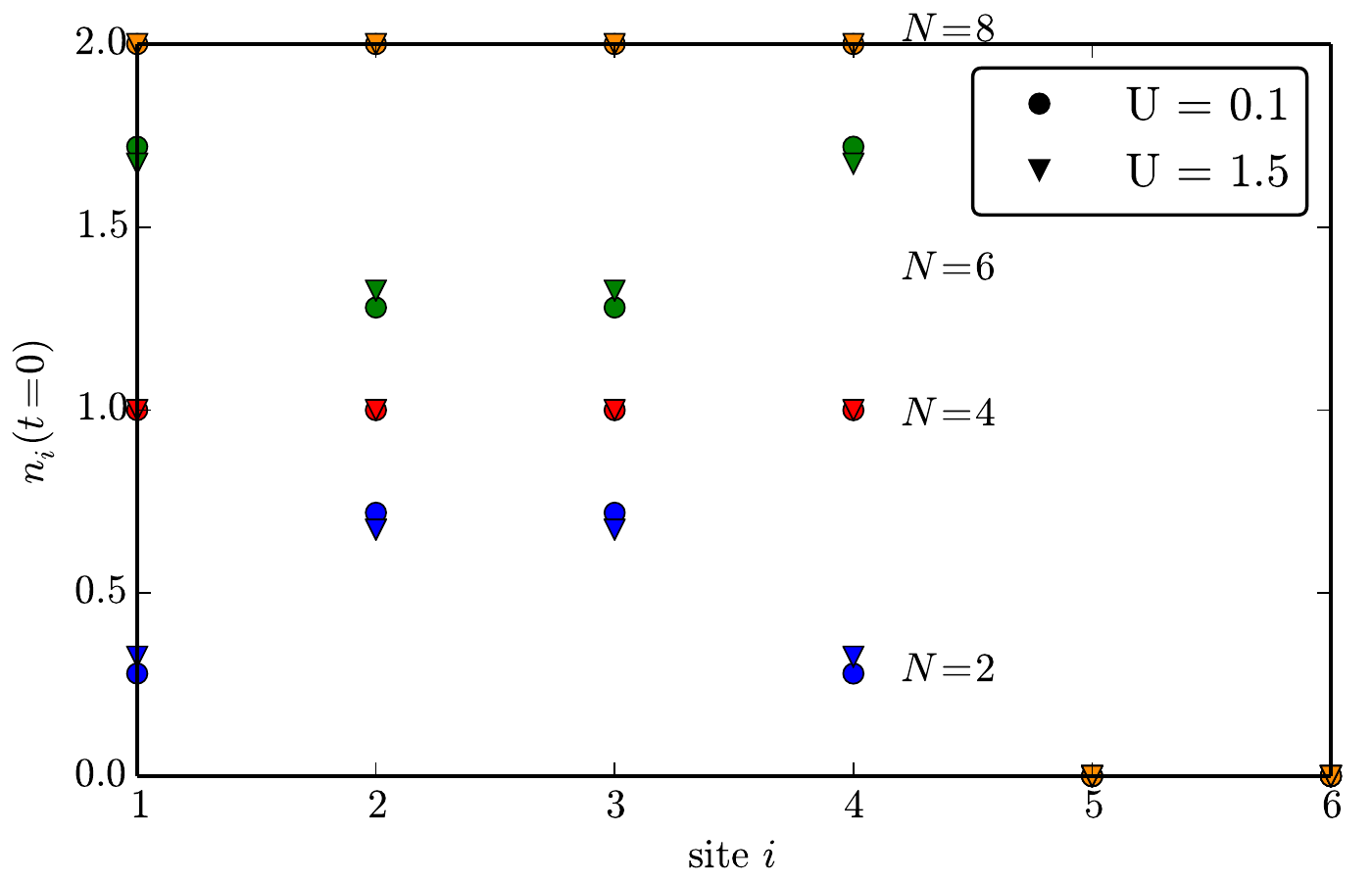}\\[-3.5ex]
\label{fig:density_t0}
\end{SCfigure}
\section{Numerical results and discussion}\label{s:results}
\begin{SCfigure}[\sidecaptionrelwidth][t]
 \includegraphics[width=0.7\textwidth]{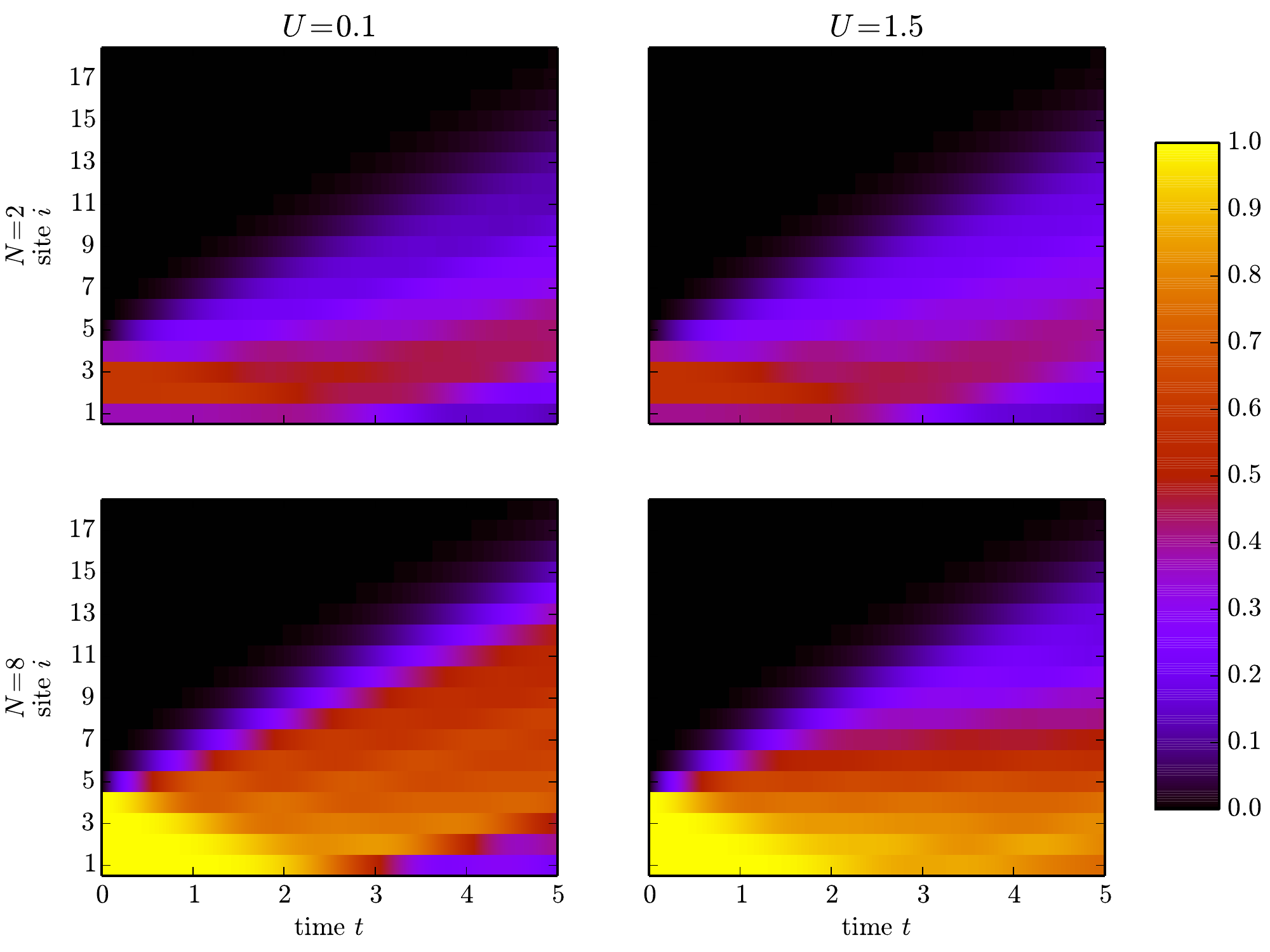}\\[-3.5ex]
% \vspace*{-1cm}
 \caption{Density evolution for $N=2$ (top row) and $N=8$ (bottom) particles for weak (left column) and strong (right) coupling starting from the initial state shown in Fig.~\ref{fig:density_t0}.}
 \label{fig:density_overview}
\end{SCfigure}

We now apply our NEGF approach with the T-matrix selfenergy (\ref{eq:sigma_t}) combined with the HF-GKBA (\ref{gkbadef}, \ref{eq:hf_prop}) to the Hubbard model (\ref{eq:h0}). To eliminate unphysical short-time dynamics, the initial state has to be chosen consistently. Here, we use a Hartree-Fock state\footnote{The influence of initial correlations, e.g., \cite{semkat_jmp00,stefanucci13} will be analyzed in a forthcoming paper} which gives rise to a slightly inhomogeneous density distribution $n_i(t=0)$ across the occupied sites, cf. Fig.~\ref{fig:density_t0}. This is a finite size effect that is particularly strong away from (initial) half ($N=4$) or full ($N=8$) filling and will be further reduced with increasing $N$. Note that the density modulation is only 
weakly affected by the pair interaction which can be seen by comparing the curves for $U=0.1$ and $U=1.5$; for increasing $U$ the density becomes flatter. Starting from this initial state, we follow the correlated dynamics of the electrons that follows after rapid removal of the virtual barrier between sites 4 and 5. The overall evolution is shown in Fig.~\ref{fig:density_overview} for two particle numbers, $N=2, 8$ and weak ($U=0.1$) and moderate ($U=1.5$) coupling. First one notices that, for $N=2$, coupling has only very little influence on the density spread. In contrast, for $N=8$ the evolution is much faster in the case of weak interaction. These trends are analyzed more in detail in Figures \ref{fig:n-dependence} and \ref{fig:u-dependence} below.

Many methods exist to quantify particle transport ranging from computation of the velocity autocorrelation function (linear response theory) to the mean square displacement $u$ (MSD), e.g., \cite{ott_prl09}. For the present study, we have to avoid any a-priori assumption of linear response or local equilibrium, so a comparison with the MSD is more appropriate. In classical systems in thermodynamic equilibrium, $u=\langle [r(t)-r(0)]^2\rangle^{1/2} ~ \sim D t^{\alpha}$, where $D$ is the diffusion coefficient and $\alpha$ the diffusion exponent which equals $1$ for ``normal'' diffusion consistent with Fick's law. Deviations from normal diffusion are well known, in particular in 2D systems \cite{ott_pre08} and may occur in the course of the relaxation \cite{ott_prl09}. For quantum systems on a lattice, $u$ is not directly accessible but, instead, one can study the time-dependent spreading of the diameter $d$ of the whole density cloud \cite{ronzheimer13,schneider_np12}, $d(t)=\sqrt{R^2(t)-R^2(0)}$, where the cloud radius $R$ (in units of the lattice spacing) is computed from the NEGF according to $R^2(t)=N^{-1}\sum_{i=1}^{N_s} n_i(t)\left(i-l_0(t)\right)^2$ where $l_0(t)=N^{-1}\sum_{i=1}^{N_s} n_i(t) i\,$ is the position of the density center, which always starts at $l_0(0)=2.5$. We also study the temporal change of the diameter, i.e. the expansion velocity, $v(t)=\frac{d}{dt} d(t)$. 
\begin{SCfigure}[\sidecaptionrelwidth][t]
 \includegraphics[width=0.6\textwidth]{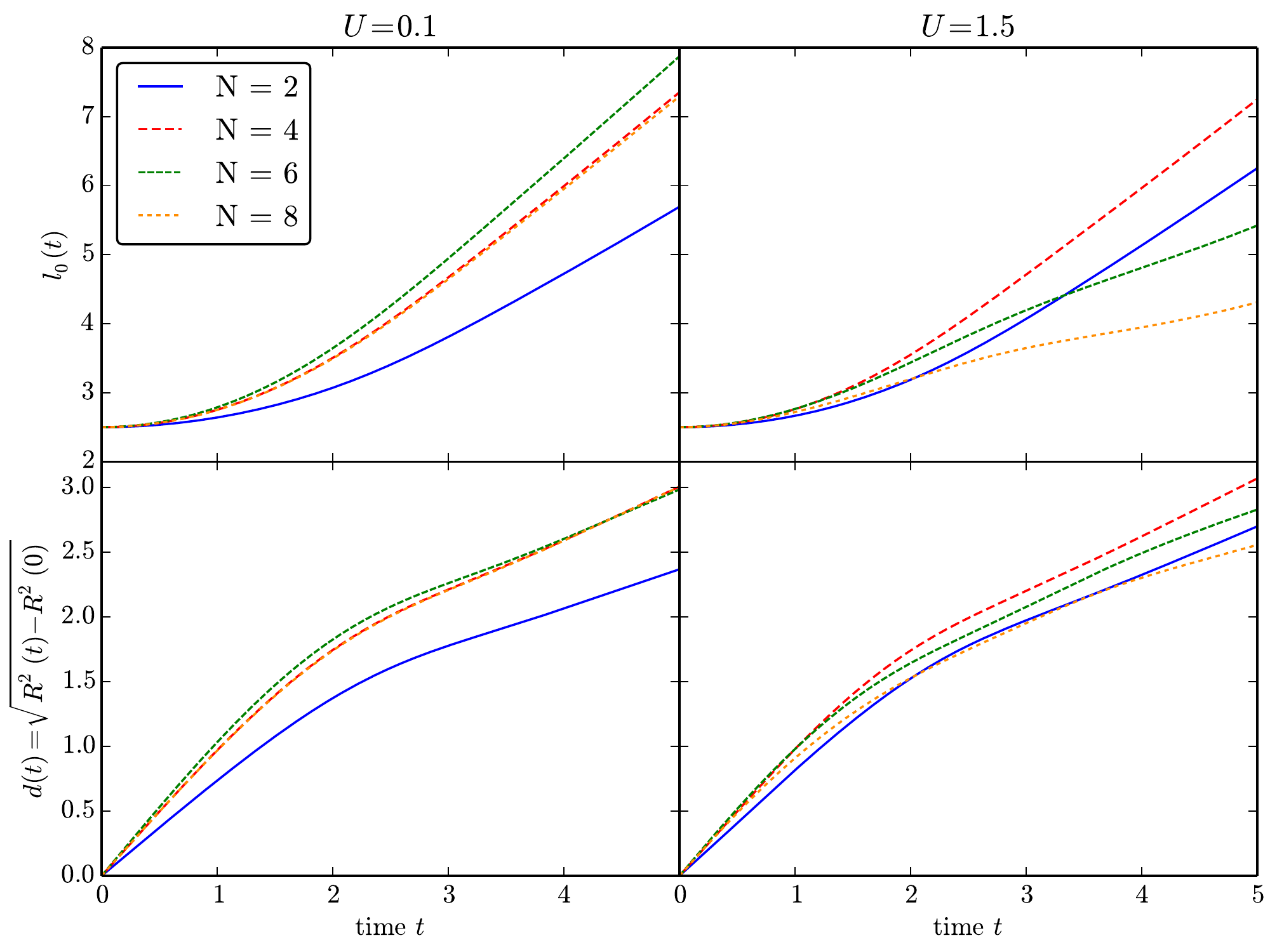}\\[-3.5ex]
 \caption{Time-dependence of the density cloud for 4 different particle numbers and weak (left column) and strong (right) coupling for the initial density shown in Fig.~\ref{fig:density_t0}. {\bf Top}: density peak position (averaged over all sites). {\bf Bottom}: width of density cloud.}
 \label{fig:n-dependence}
\end{SCfigure}

In Fig.~\ref{fig:n-dependence}, we show the time dependence of the cloud position, $l_0(t)$, and width, $d(t)$, on the particle number. For weak coupling the center of mass (com) velocity of the cloud increases from $N=2$ to $N=6$ and decreases again for $N=8$.
Similarly, the cloud width increases slowest for $N=2$ whereas there is almost no change for $N=4\dots 8$. For moderate coupling, $U=1.5$, the picture is different. While the com velocity again increases from $N=2$ to $N=4$, it is strongly reduced for $N=6$ and $N=8$. We now turn to our main topic---the dependence of particle transport on the coupling strength $U$ which is analyzed in Fig.~\ref{fig:u-dependence} for the two cases $N=2$ and $N=8$. As noticed in Fig.~\ref{fig:n-dependence} before, the cases of 2 particles exhibits only very weak dependence on $U$. In contrast, for $N=8$ some trends are obvious. The com velocity decreases monotonically with $U$ (top right Fig.), and also the width grows slower when $U$ is increased (middle right Fig., except for $U=0.1 \dots 1$).
Finally, let us consider the expansion velocity $v(t)$ of the cloud (bottom row). Considering the horizontal asymptotics of $v$, a few trends seem to be noticeable. For $N=2$ the expansion velocity (and with it the diffusion coefficient) increases monotonically with $U$. For $N=8$ the trend is less clear, although an overall reduction with $U$ seems to occur, but the simulations are still too short for a definite conclusion about the monotonicity of this trend. We note that in experiments with fermionic atoms in an optical lattice, a monotonic reduction was observed for $U=0\dots 2$ after which the velocity increased again \cite{ronzheimer13}.
\begin{SCfigure}[\sidecaptionrelwidth][t]
 \includegraphics[width=0.5\textwidth]{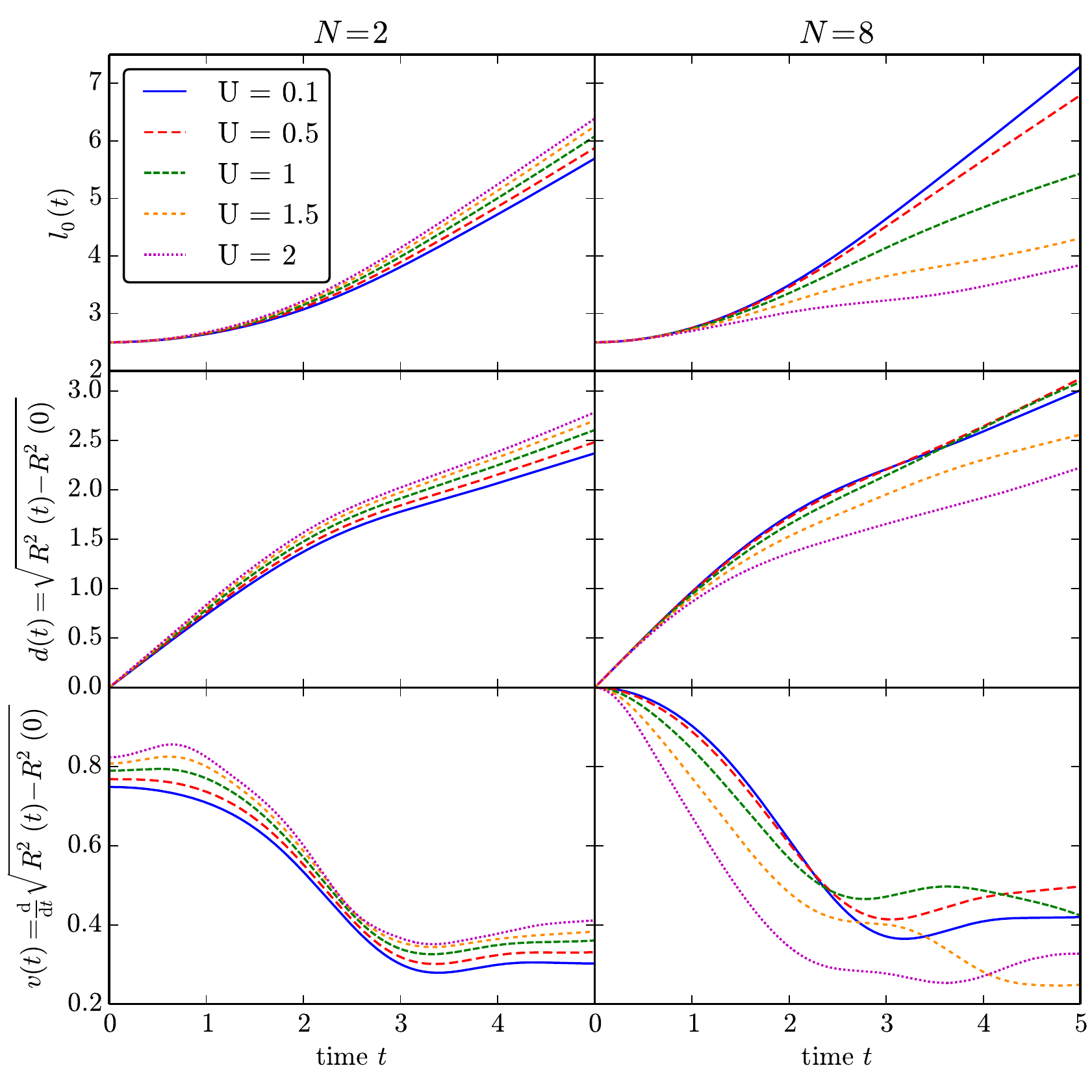}\\[-5.5ex]
 \caption{Time-dependence of the density cloud for the cases of $N=2$ particles (left column) and $N=8$ particles (right) and five different values of the coupling strength. {\bf Top row}: density peak position (averaged over all sites). {\bf Center row}: width of density cloud. {\bf Bottom}: Time derivative of the width. If the evolution approaches a horizontal curve the asymptotic value is related to the diffusion coefficient \cite{ronzheimer13}.}
 \label{fig:u-dependence}
\end{SCfigure}

To summarize, despite the preliminary character of our results, we have demonstrated that the present NEGF approach with T-matrix selfenergies combined with the HF-GKBA is able to tackle the problem of strongly correlated fermions on a lattice \cite{puigvonfriesen09}. The simulations yield the full time dynamics under fully inhomogeneous conditions far from equilibrium. Our results for small systems do well agree with exact diagonalization results \cite{hermanns_prb14} which, however, are not feasible for the present lattice dimension. While the problem of correlated particle transport can be studied with high accuracy using time-dependent density matrix renormalization group methods \cite{ronzheimer13}, these simulations are restricted to 1D systems. We also mention simulations based on a classical kinetic equation in relaxation time approximation \cite{schneider_np12}, but here fundamental questions regarding conservation laws remain open.
The present NEGF approach should be able to overcome these limitations. It is total energy conserving and applicable to systems of arbitrary dimensionality. We are presently extending the simulations to larger systems and longer times as well as to larger $U$ so that the problem of diffusion will be solved in a more quantitative way, taking into account strong correlation effects such as doublon formation \cite{kajala_prl11, schluenzen14}.
\begin{acknowledgement}
 This work was supported by the Deutsche Forschung Gemeinschaft via grant BO1366-9 and the Northern German Supercomputing Alliance (HLRN) via grant shp006.
\end{acknowledgement}

%----------------------

\end{document}